\documentclass[prl,aps,amsmath,amssymb,amsfonts,floatfix,groupeaddress,showpacs,twocolumn]{revtex4-1}
\usepackage{graphicx}
\usepackage[utf8]{inputenc}
\usepackage{epsfig}
\usepackage[dvips]{color}
\usepackage{nicefrac}
\usepackage{float}
\usepackage{tabularx}
\newcolumntype{L}[1]{>{\raggedright\arraybackslash}p{#1}} 
\newcolumntype{C}[1]{>{\centering\arraybackslash}p{#1}} 
\newcolumntype{R}[1]{>{\raggedleft\arraybackslash}p{#1}} 
\newcommand{\abs}[1]{\left| #1 \right|}
\newcommand{\ham}{{\cal H}}
\newcommand{\bk}{\mathbf{k}}
\newcommand{\matrixelement}[3]{\left< #1 \left | #2 \right | #3 \right>}
\newcommand{\cdaggop}[1]{c_{#1}^{\dagger}}
\begin{document}

\title{Large-amplitude spin oscillations triggered by nonequilibrium strongly correlated $t_{2g}$ electrons}
\author{Malte Behrmann}
\affiliation{I. Institut f{\"u}r Theoretische Physik,
Universit{\"a}t Hamburg, 20355 Hamburg, Germany}
\author{Frank Lechermann}
\affiliation{I. Institut f{\"u}r Theoretische Physik,
Universit{\"a}t Hamburg, 20355 Hamburg, Germany}

\begin{abstract}
Laser-induced ultrafast (fs) magnetization experiments in antiferromagnets have recently attracted large attention, paving the road for inherently fast spin dynamics in the THz regime without invoking stray fields. The technical importance is emphasized by the rising new research field of antiferromagnetic (AFM) spintronics, where superexchange-dominated strongly correlated compounds provide an interesting materials playground. An intriguing question is whether the Coulomb interaction may be a key to control AFM order on ultrafast time scales. Therefore, we study (de)magnetization processes in a time-dependent multiorbital Hubbard model, focusing on $t_{2g}$ electrons in a wider doping range. Depending on filling, we reveal large-amplitude spin oscillations via interaction quenches from the antiferromagnetic or paramagnetic state. Nonequilibrium ultrafast spin-orientation effects in prominent correlated transition-metal oxides are therefrom predicted.  
\end{abstract}

\maketitle
\section{I. Introduction}
Since the discovery of ultrafast demagnetization of ferromagnetic (FM) Ni by a 60-fs laser pulse~\cite{beaurepaire_ultrafast_1996}, multiple experimental studies have studied similar magnetization dynamics in ferromagnets~\cite{kimel_laser-induced_2004,kimel_ultrafast_2005,kimel_inertia-driven_2009,vahaplar_ultrafast_2009}. A focus lies on switching spin orientations in a deterministic way on shortest (femtosecond) time scales. This is critical for data storage as it sets the bit-recording time limit in magnetic memory devices~\cite{pan_data_2009}. Many theoretical mechanisms have been proposed to explain the ultrafast demagnetization and a way to control FM order. These include the Elliot-Yafet mechanism~\cite{steiauf_elliott-yafet_2009,koopmans_explaining_2010}, superdiffusive spin transport~\cite{battiato_superdiffusive_2010} and processes driven by the Coulomb interaction~\cite{kraus_ultrafast_2009}. 

The latter plays the dominant role in equilibrium strongly correlated materials, often leading to antiferromagnetic (AFM) order via superexchange. Notably for compounds harboring transition-metal ions with $3d$ shells close to half filling. In this context, the spintronics of AFM systems is a rising focus in the research on nonequilibrium systems~\cite{gomonay_spintronics_2014,nunez_theory_2006,duine_inelastic_2007,haney_ab_2007}. It offers inherently faster processes with spin dynamics in the THz regime~\cite{sievers_far_1963,satoh_spin_2010,nishitani_terahertz_2013}. Such has been investigated in the fully compensated ($S^{\rm (tot)}=0$), strongly correlated antiferromagnets NiO~\cite{satoh_spin_2010,lefkidis_angular_2009,lefkidis_first-principles_2013} and MnO \cite{nishitani_terahertz_2013} as well as the rare-earth orthoferrites~\cite{kimel_laser-induced_2004,kimel_ultrafast_2005,kimel_inertia-driven_2009,jiang_dynamical_2013}. 
Finally, the exchange-bias effect~\cite{meiklejohn_new_1956} can in principle be used to employ ultrafast control of FM order by manipulating the magnetic state of an adjacent antiferromagnet~\cite{le_guyader_dynamics_2013}). It is important to know whether the AFM ordering especially in correlated materials can be controlled via general mechanisms, not readily accessible by weak-coupling approaches.

Here, we indeed show that Coulomb interactions in a multiorbital system lead upon laser excitation to large-amplitude oscillations of the AFM order parameter. This behavior is evoked by tuning the initial magnetic ground state as well as the excitation strength. With the generic model study we aim at a broad materials class of correlated transition-metal oxides with dominant $t_{2g}$ physics from $N$ electrons. For instance, high-N{\'e}el-temperature SrTcO$_3$~\cite{rodriguez_high_2011} (half-filled $N=3$), nearly-AFM SrCrO$_3$~\cite{ortega-san-martin_microstrain_2007} ($N=2$) and paramagnetic (PM) Sr$_2$MoO$_4$~\cite{ikeda_magnetic_2000} ($N=2$). These $t_{2g}$ materials are subject to peculiar effects of strong correlations, driven not only by an Hubbard $U$ but notably a relevant Hund's exchange $J_{\rm H}$~\cite{de_medici_janus-faced_2011,rodriguez_high_2011,werner_spin_2008,mravlje_origin_2012,de_medici_janus-faced_2011}. Using a generic multiband Hubbard model, we show that those Coulomb interactions are also the key players concerning nonequilibrium magnetism on the few-hundred fs time scale. As we are aiming here for general implications with materials-realistic fillings, an orbital-degenerate study is performed. Note that experiments in fact put SrCrO$_3$ and Sr$_2$MoO$_4$ in this category of vanishing crystal field within the $t_{2g}$ manifold~\cite{ortega-san-martin_microstrain_2007,ikeda_orbital-degenerate_2000}. Our results reveal an inherent connection between (transient) oscillatory behavior and the proximity of the excited state to an equilibrium AFM-PM phase boundary.
\section{II. Theoretical Framework}
This study focuses on interacting $t_{2g}$ electrons within a model context. Our three-band Hubbard Hamiltonian on the simple-cubic lattice uses a nearest-neighbor (NN) hopping $t_h$ and full rotational-invariant Coulomb interactions in Slater-Kanamori parameterization, i.e. utilizing Hubbard $U$ and Hund's exchange $J_{\rm H}$. In the following the half-bandwidth $D$ sets the energy scale. The complete Hamiltonian $\ham$ reads as 
\begin{equation}
\ham = -t_h \sum_{\langle i,j\rangle p\sigma}\left(\cdaggop{ip\sigma} c^{\hfill}_{jp\sigma}+{\rm h. c.}\right) + \sum_i \ham_i^{\rm loc}\;\;,\label{eq:ham}
\end{equation}
with the interaction term on each site written as
\begin{eqnarray}
\ham_i^{\rm loc}&=& U\sum_{p} n_{p\uparrow}n_{p\downarrow}+\nonumber\\
&&\hspace*{-0.5cm}+\frac 12 \sum \limits _{p \ne p',\sigma}\hspace*{-0.3cm}
\Big\{\left(U\hspace*{-0.1cm}-\hspace*{-0.1cm}2J_{\rm H}\right) \, n_{p \sigma} n_{p' \bar \sigma}
+ \left(U\hspace*{-0.1cm}-\hspace*{-0.1cm}3J_{\rm H}\right) \,n_{p \sigma}n_{p' \sigma}\nonumber\\
&&\hspace*{-0.5cm}+\left.J_{\rm H}\left(c^\dagger_{p \sigma} c^\dagger_{p' \bar\sigma}
c^{\hfill}_{p \bar \sigma} c^{\hfill}_{p' \sigma}
+c^\dagger_{p \sigma} c^\dagger_{p \bar \sigma}
 c^{\hfill}_{p' \bar \sigma} c^{\hfill}_{p' \sigma}\right)\right\}\;,\\
 &&\hspace*{-0.75cm}= \left(U\hspace*{-0.1cm}-\hspace*{-0.1cm}3J_{\rm H}\right) \frac{\hat{N}(\hat{N}-1)}{2}+\frac{5}{2}J_{\rm H}\hat{N} -2J_{\rm H} \vec{S}^2-\frac{1}{2}J_{\rm H}\vec{L}^2\;. \nonumber
\end{eqnarray}
Here, the indices $p,p'=1$,2,3 are labeling $t_{2g}$ orbitals and $\sigma=\uparrow$,$\downarrow$
is the spin projection. The quantities $\hat{N}$, $\vec{S}$ and $\vec{L}$ mark the particle, spin and angular-momentum operator. This renders the local symmetries $L^2$, $S^2$ and $S_z$ obvious in the given model representation. 

The equilibrium interacting problem is solved within rotational-invariant slave-boson mean-field 
theory (SBMFT)~\cite{li_spin-rotation-invariant_1989,lechermann_rotationally_2007},
with the same level of approximation as a corresponding Gutzwiller formulation~\cite{bue98}.
Out of equilibrium the time-dependent (TD) extension to SBMFT is applied~\cite{schiro_time-dependent_2010,behrmann_extended_2013}, which is tailored to the short-time dynamics with and without magnetic order~\cite{schiro_time-dependent_2010,sandri_nonequilibrium_2013,bunemann_linear-response_2013}. It may address the correlated metallic as well as the Mott-insulating state in the nonequilibrium regime. Qualitatively, the same physics as more advanced full Keldysh-contour 
schemes~\cite{eckstein_thermalization_2009,tsuji_nonthermal_2013} is reproduced, aside from 
general aspects of thermalization. 

For the dynamic regime, an SBMFT determined equilibrium solution from free-energy 
considerations sets the stage. Afterwards, we propagate the equilibrium solution with a TD 
Hamiltonian. The condensed slave bosons $\phi$ become time dependent and the following set 
of nonlinear differential equations (site index $i$ suppressed) is numerically solved:
 \begin{eqnarray}
 \imath \frac{\partial \nu_{a\alpha}^\bk}{\partial t} &=&
 \sum_\beta \tilde{H}_{\alpha \beta}^\bk
 \nu_{a \beta}^\bk \label{td_dgl1}\;,\;\;\;
 \tilde{H}_{\alpha \beta}^\bk =
 \sum_{\alpha' \beta'} R^\dagger_{\alpha \alpha'}
  \varepsilon^{\bk}_{\alpha' \beta'} R_{\beta'\beta}\;,\quad\\
 \imath \frac{\partial \phi_{AB}}{\partial t} &=& \sum_C \ham^{\rm loc}_{AC} \phi_{CB} +
 \sum_{\bk b}^{\rm occ}\sum_{\alpha \beta} \nu_{b \alpha}^{*\bk}
 \frac{\partial\tilde{H}_{\alpha \beta}^\bk} {\partial \phi_{AB}^\dagger}\nu_{b \beta}^\bk
 \label{td_dgl2}\;.
 \end{eqnarray}
 In these equations, $a,b$ label eigenvalues, $\nu^\bk$ eigenstates of
 $\tilde{H}[\phi]$ in momentum ($\bk$) space, $A,B,C$ denote the local basis states and $\alpha,\beta$ the orbital-spin combination. The dispersion $\varepsilon^{\bk}$ results
from the Fourier transform of the NN-hopping kinetic term of eq.~(\ref{eq:ham}).
A numerical solution of eqns.~(\ref{td_dgl1},\ref{td_dgl2}) is achieved by using an adaptive Runge-Kutta scheme of order 6(5)~\cite{verner_numerically_2010}.

The laser excitation (sufficiently short regarding pulse width) is modeled by an interaction quench. This is realized by choosing an initial $U_i$($t=0$) and a final $U_f$($t>0$). The ratio $q \equiv J_{\rm H}$/$U$ is kept fixed, i.e. $q=q_i=q_f$. Surely, an interaction quench does not cover all the details of a
realistic laser-excitation process, and other approximative TD Hamiltonians are conceivable. But, the present choice is suitable
to reveal the key effects occuring in the short-time regime of correlated multiorbital magnetization dynamics. Our filling $N$ is defined as $t_{2g}$ electrons per site in a two-site unit cell, i.e. $N=3$ marks half-filling. Due to the absence of crystal-field splitting, the magnetic moment $m$ is orbitally degenerate and here given per orbital and per site in units of $2\mu_B$. Note that recent work by Sandri {\sl et al.}~\cite{sandri_nonequilibrium_2013} showed that the present formalism addresses the AFM-to-PM transition in an interaction quench of the one-band Hubbard model qualitatively correct compared to more elaborate schemes~\cite{tsuji_nonthermal_2013}. This provides strong confidence that the method performs also well for the more challenging multiorbital problem at hand.
 \section{III. Results}
\begin{figure}[b]
\hspace*{-0.35cm}
\includegraphics*[width=8.75cm]{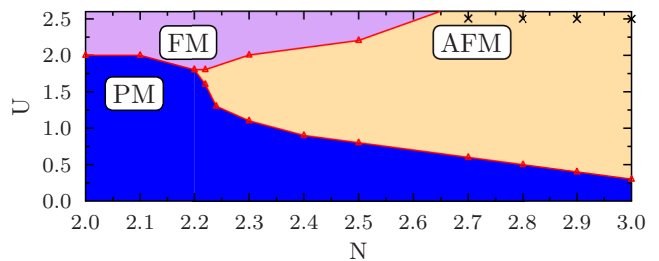}
\caption{(Color online) Equilibrium SBMFT magnetic phase diagram for the hole-doped three-band Hubbard model on the cubic lattice with $q$=0.2. Black crosses mark for comparison stable 
AFM solutions within DMFT using an CT-QMC impurity solver~\cite{gull_continuous-time_2011,boehnke_orthogonal_2011,parcollet_triqs_????} at $\beta t_h$=50 and $q=0.167U$ (see text).\label{static_phasediagram}}
\end{figure}
    \begin{figure}[t]
    \includegraphics*[width=8.75cm ]{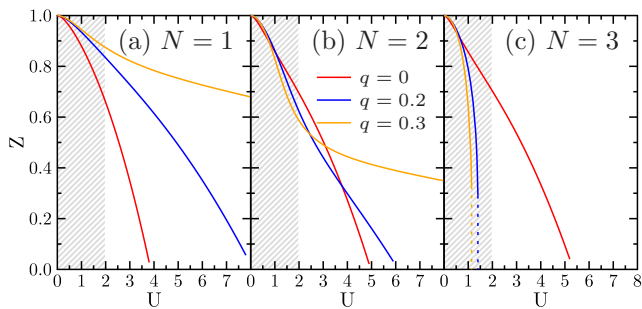}
      \caption{(Color online) Equilibrium quasiparticle weight $Z=(1-\frac{\partial\Sigma}{\partial\omega}|_{\omega\rightarrow 0})^{-1}$ for different $U$, $q$, and fillings $N=$1,2,3. Grey shaded areas mark the interaction regime covered by the nonequilibrium study in the subsequent quench scenarios.}
      \label{eq_zmatrix}
    \end{figure}
 \subsection{A. Equilibrium case}
Before entering the nonequilibrium regime, we recap relevant characteristics of the equilibrium problem. The magnetic phase diagram with hole doping from SBMFT for $q=0.2$ shown in Fig.~\ref{static_phasediagram} displays a PM phase for small $U$ and an AFM/FM phase at lower/higher doping for larger $U$. Within dynamical mean-field theory (DMFT) using a continous-time quantum Monte Carlo (CT-QMC) solver, Chan {\sl et. al}~\cite{chan_magnetism_2009} found no AFM order away from half filling on the Bethe lattice for $q$=0.167 at $\beta t_h$=50 ($\beta$ is inverse temperature). But, our comparative DMFT computations with same parameter setting on the cubic lattice reveals indeed stable antiferromagnetism for moderate doping~(see Appendix).

It is known that the impact of the Hund's exchange is a key feature for PM ground states in the doped three-band Hubbard model~\cite{de_medici_janus-faced_2011}. In Fig.~\ref{eq_zmatrix} this so-called {\sl Janus-faced} influence is clearly seen for the filling $N=2$. 
There, for larger $J_{\rm H}$ the correlation strength is increased at smaller $U$, but the
Mott transition is shifted to much larger $U$. However, note that in this work the nonequilibrium study is restricted to $U_f\leq 2.0$, as this is the equilibrium interaction region where the AFM ground state is stable over a broad region of filling $N$ ($2.2 \leq N \leq 3.0$). Thus, there is no strong Janus-faced influence of $J_{\rm H}$ onto the nonequilibrium magnetic responses expected. The detailed impact of the Janus-faced physics on itinerant magnetically ordered states is still an open question.

\subsection{B. Nonequilibrium case\label{sec:noneq}}
\begin{figure}[t]
 \includegraphics*[width=8.5cm]{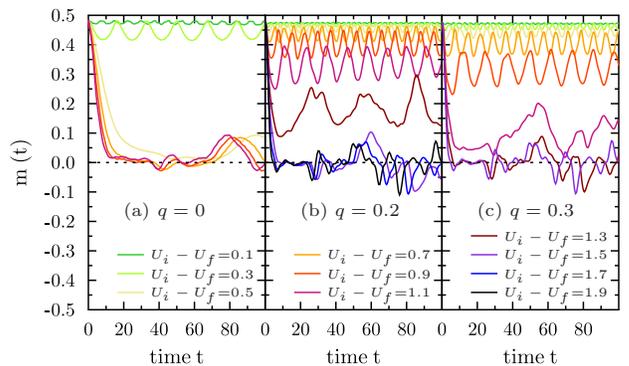}
 \caption{(Color online) TD magnetization for different $q$ at half-filling after 
 the quench. The chosen $U_i$ ensure equal initial magnetization: 
 (a) $U_i=3$, (b) $U_i=2$, (c) $U_i=1.6$. }
 \label{N3_scenario1_magn}
 \end{figure}

Before considering the doping-dependent scenario, we provide a connection to earlier single-band studies~\cite{tsuji_nonthermal_2013,sandri_nonequilibrium_2013} by exploring the dynamic magnetic response at half-filling from an initial AFM state towards a nonequilibrium PM state at smaller interaction values. This allows to investigate the influence of different $J_{\rm H}=qU$ onto the magnetization dynamics. Figure~\ref{N3_scenario1_magn} shows the time evolution of the magnetic moment for a single site with equal initial $m$ and different $J_{\rm H}$ values. The quantity $U_{fc}^{\rm PM}$ marks the critical final interaction value needed to observe PM behavior. A vanishing Hund's exchange massively decreases the interaction difference $U_i-$$U_{fc}^{\rm PM}$. As the AFM correlations weaken with doping, $U_i$$-$$U_{fc}^{PM}$ is largest at half-filling. In the following, we choose $q=0.2$, but qualitative changes induced by increasing $q$ are also discussed.

\begin{figure}[t]
\centering
\includegraphics*[width=8.7cm]{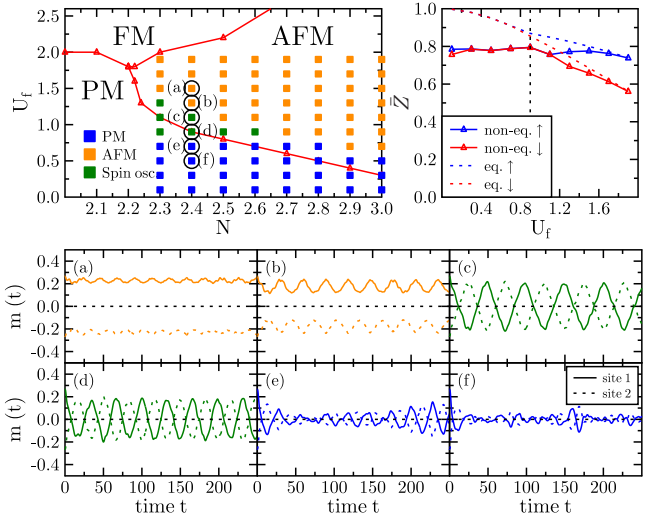}
  \caption{(Color online) Top left: Nonequilibrium demagnetization response diagram for $q=0.2$. Full lines separate the equilibrium phases at $U=U_f$. Initial AFM state is established for $U_i=2$ and all $N$. Squares indicate $U_f$ and the corresponding nonequilibrium magnetic response. Top right: Time-averaged ($\bar{Z}$) and equilibrium QP weight at $N=2.4$. Bottom: TD magnetization $m(t)$ for selected $U_f$ at $N=2.4$.}
  \label{phasediagram_scenario1}
\end{figure}

\textit{Initial AFM state}.---
The first hole-doped scenario considers the dynamic demagnetization from an initial metallic AFM state (at $U_i=2$) to a final metallic PM state at small $U_f$, keeping $q=0.2$. As exemplified for $N=2.4$ (see Fig.~\ref{phasediagram_scenario1}) small quench strengths lead to a dynamic AFM state with small variation of the staggered magnetic moment. Reducing $U_f$ further triggers a new kind of magnetic response. A spin oscillation with large amplitude around the zero staggered moment sets in, leading to a periodic sign change of the magnetic moment on each site. This spin oscillation exhibits still an antiferromagnetic order concerning neighboring sites in the unit cell (site 1, site 2, see Fig.~\ref{phasediagram_scenario1}). Note that the time-averaged quasiparticle (QP) weight $\bar{Z}$ is lower than the corresponding equilibrium $Z$. For even smaller $U_f$, the PM state is finally reached and the QP weight becomes nearly $U_f$ independent with a significantly lower value than the equilibrium one. Taking into account the variation of the filling $N$ in Fig.~\ref{phasediagram_scenario1} renders it clear that the non-equilibrium AFM-PM response boundary is indeed rather close to the corresponding equilibrium phase boundary. With doping away from half-filling, the intriguing AFM spin oscillations start to appear at $N$$\sim$2.6. The susceptible $U_f$ range for these oscillations
broadens towards $N\sim2.2$, where the equilibrium AFM phase eventually breaks down.
It seems likely that the occurrence of these large-amplitude spin oscillations is connected to enhanced magnetic 
fluctuations close to the phase boundary. Some features are reminiscent of behavior close to a nonthermal critical 
point, which has been previously found in time-dependent single-orbital studies~\cite{tsuji_nonthermal_2013,sandri_nonequilibrium_2013}. There, the frequency of the TD magnetic moment is known to tend to zero from above. From the bottom graphs in Fig.~\ref{phasediagram_scenario1}, the main frequency of the changing local magnetic moment shows a similar characteristic. It decreases from $U_f$=1.5-1.1 and then increases again at $U_f$=0.9. Note that thermalization is delayed near such a (nonthermal) critical point in the single-orbital case~\cite{tsuji_nonthermal_2013}.

To better understand the origin of the large-amplitude spin oscillations, examining the occupations of the dominant local multiplets is insightful. Those are here given by the ($L=S=1$) spin triplet ($\phi_{L_z,1,S_z}$) and the ($L=0$, $S=\frac{3}{2}$) spin quartet ($\phi_{\frac{3}{2},S_z}$) (see Table~\ref{tab_local_states_scenario1}). Both are expressed through the eigenvalues of the squared angular momentum and spin operator, i.e. $\hat{L}^2$ and $\hat{S}^2$ (see also Ref.~\cite{de_medici_janus-faced_2011}). One thus can define a threshold parameter $\eta$ via ratios between the maximal amplitude of the TD magnetic 
fluctuations for the spin triplet/quartet and the initial (equilibrium) spin polarization, i.e.
\begin{eqnarray}
\label{spin_osc_threshold_formula}
\eta = \frac{w_1(A^1_1-A^1_0)}{\abs{\phi^{(t=0)}_{1,1}}^2-\abs{\phi^{(t=0)}_{1,0}}^2}+
\frac{w_\frac{3}{2}\left(A^\frac{3}{2}_\frac{3}{2}+A^\frac{3}{2}_{-\frac{3}{2}}\right)/2}
{\abs{\phi^{(t=0)}_{\frac{3}{2},\frac{3}{2}}}^2
-\abs{\phi^{(t=0)}_{\frac{3}{2},-\frac{3}{2}}}^2}\,, \hspace{0.5cm} \\
w_S=\sum_{S_z=\pm S,\pm S -1} \abs{\phi^{(t=0)}_{S,S_z}}^2\,,\;A^S_{S_z} = {\rm Max}\left(\abs{\phi^{(t)}_{S,S_z}}^2\right).\;\nonumber
\end{eqnarray} 
\begin{table}[b]
\begin{tabular}{  l | c  c  c  c}
  $\{L,S\}$ ($S_z$) & $N$=2.3 &  $N$=2.5 &  $N$=2.8 &  $N$=3.0 \\[0.05cm]
  \hline\\[-0.25cm]
  $\{1,1\}$ (1)  & 0.50 (0.32)  & 0.41 (0.35)  & 0.22 (0.21) & 0.05 (0.05) \\[0.05cm]
  \hline\\[-0.25cm]
  $\{0,\frac{3}{2}\}$ $(\frac{3}{2})$  & 0.31 (0.17)  & 0.44 (0.35)  
  & 0.69 (0.66) & 0.88 (0.88)\\ 
  \end{tabular}
\caption{Initial occupation at $t=0$ of selected single-site multiplets for different fillings $N$ and $U=2$, $q=0.2$.}\label{tab_local_states_scenario1}
\end{table} %
As the spin triplet is $L_z$ degenerate, we understand $\phi_{1,S_z}$=$\sum_{L_z=\pm 1,0} \phi_{L_z,1,S_z}$. The maximum amplitudes $A$ are computed from the maximum value (Max) of the slave bosons in the time interval $[10,250]$. The first term in eq.~(\ref{spin_osc_threshold_formula}) arises from the difference in $A$ between $S_z=1$ and $S_z=0$ of the triplet. As the quartet has no $S_z=0$ state, the second term originates from the average amplitude of states with largest $S_z$ difference, since those are most susceptible to magnetic fluctuations. In order to normalize the different filling scenarios, both contributions are weighted with the initial multiplet occupation. Note that the inspected time interval starts beyond the initial drop in $m$ from dephasing. The TD magnetic fluctuations overcome the initial spin polarization for $\eta>0.5$ and thus can invert the magnetization (see Fig.~\ref{phaseobs_scenario1}) (for a consideration of the energetics, see Appendix). Close to half-filling, the increased initial spin polarization of the spin quartet (see Table~\ref{tab_local_states_scenario1}) dominates the dynamic magnetic amplitude and the large-amplitude spin oscillations disappear. Small shifts of the latter region of appearance in $U_f$ occur by an increased $q=0.3$. But the filling dependence remains qualitative identical. Thus, the Hund's $J_{\rm H}$ coupled three-particle quartet and two-particle triplet rule the doped $t_{2g}$ dynamics. A break-up of these multiplets in lower $S$ excitations is not detectable on a substantial level.\\             
\begin{figure}[t]
\centering
\includegraphics*[width=8.3cm]{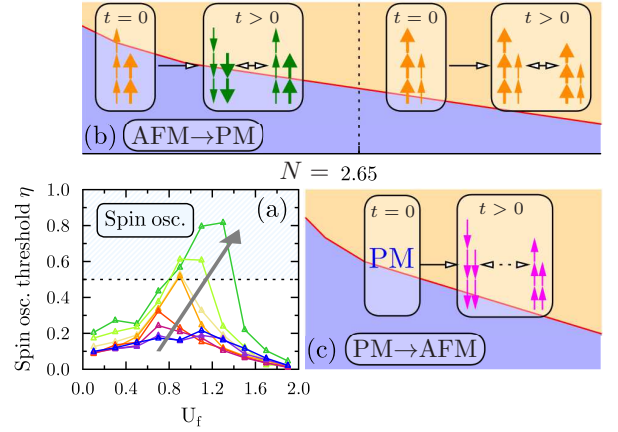}
  \caption{(Color online) (a) Spin oscillation threshold $\eta$ for various $U_f$ and 
fillings $N=3.0$,$\dots$,2.3 (from blue to green) for the quenched AFM state. (b,c) Sketched time evolution of three-particle quartet state (three arrows) and two-particle triplet state (two arrows). Green and purple colors mark spin oscillations for a quenched AFM and PM state, respectively.}
  \label{phaseobs_scenario1}
\end{figure}
\textit{Initial PM state}.---
Second, we deal with an initial metallic PM state close to the equilibrium AFM-PM phase boundary, quenched to {\sl larger} $U_f$, keeping $q=0.2$. Only initial states close to the equilibrium PM-AFM phase boundary allow for a finite TD local spin expecation value $\langle S\rangle$ upon quenching. As observable in Fig.~\ref{phasediagram_scenario2}, small quench strengths readily lead to modulated dynamic AFM order with nonequilibrium QP weight $\bar{Z}$ close to the equilibrium value. For rather large $U_f\sim$1.75$-$2 the system remains in a strongly correlated metallic PM state with $\bar{Z}<0.2$ much smaller than the corresponding equilibrium $Z$. The robust PM magnetic order for strong quenches is easily understood from the large energy transfer that raises the effective temperature above reasonable N{\'e}el scales. Interestingly, however, at intermediate quench-strength {\sl transient} AFM spin oscillations appear with a length of 20 to 100 fs~(see Appendix). They show a short-time decay into a dynamic itinerant AFM state and are substantially different to those encountered in the PM-to-AFM scenario. This new nonequilibrium feature is appearing already at low hole doping and remain vital for nearly all fillings up to the vanishing point of the static AFM phase. Moreover, there appears to be no obvious pinning to the static AFM-PM phase boundary. 
However, the noninteracting ground state as initial state with small hole doping ($N=$2.6-2.9) shifts the $U_f$ region ($U_f\sim$0.5) for transient spin oscillations right at the static AFM-PM phase boundary. The influence of thermalization onto these transient spin oscillations can be estimated from a paramagnetic single-orbital TD-DMFT study \cite{eckstein_thermalization_2009}. There, weak interaction quenches from a noninteracting initial state lead to no thermalization on intermediate time scales. This indicates that the present transient spin oscillations should be observable shortly after the quench, before thermalization sets in. 
\begin{figure}[t]
\centering
\includegraphics*[width=8.7cm]{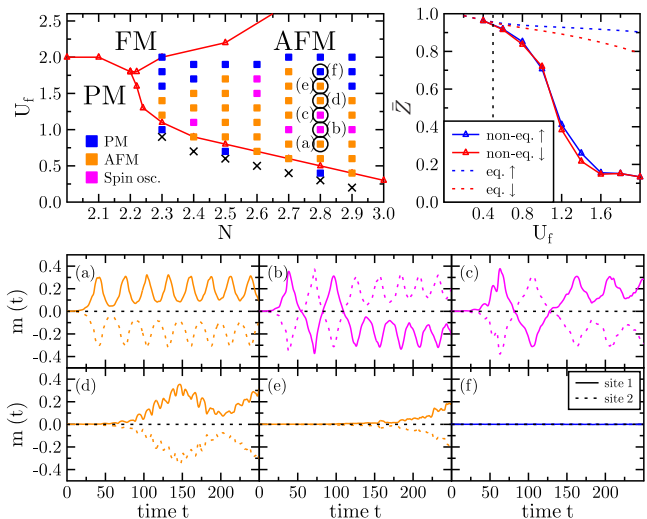}
\caption{(Color online) Top left: Nonequilibrium magnetization response diagram for $q=0.2$ with equilibrium phases separated by full lines (calculated at $U=U_f$). Crosses denote initial PM state and squares indicate $U_f$ and the corresponding nonequilibrium magnetic response. Top right: Time-averaged ($\bar{Z})$ and equilibrium QP weight at $N=2.8$. Bottom: TD magnetization $m(t)$ for selected $U_f$ at $N=2.8$.\label{phasediagram_scenario2}}
\end{figure}

\begin{figure}[t]
 \centering
 \includegraphics*[width=8.75cm]{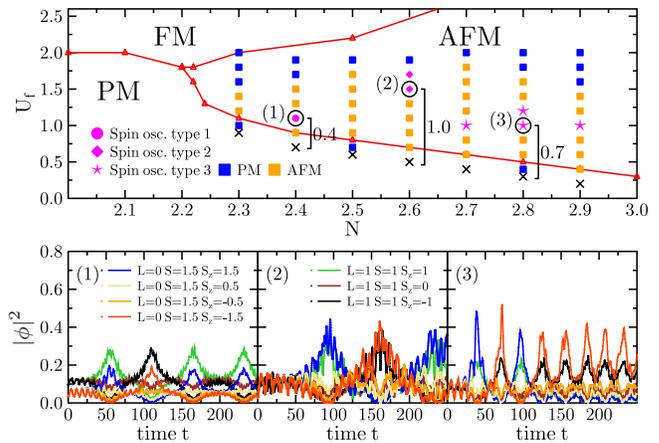}
 \caption{(Color online) Top: Distinction between three types of transient spin 
 oscillations when quenching the PM phase. Bottom: TD occupation numbers from 
 squared slave-boson amplitudes for local spin quartet and triplet within these types:
 (1) $|\phi_t|^2>|\phi_q|^2$, (2) $|\phi_t|^2\sim|\phi_q|^2$ and 
 (3) $|\phi_q|^2<|\phi_t|^2$, with $t$: triplet and $q$: quartet.
 All other markers/labels as Fig.~\ref{phasediagram_scenario2}.
 \label{sw_types_by_local_states}}
 \end{figure}

The derived threshold parameter $\eta$ of the first scenario is here not applicable, as there is no initial spin polarisation. To still shed light onto the complicated doping behavior, we consider again the same two maximally occupied sets of local states as in the demagnetization quenches. Namely, from the filling-dependent occupation hierachy between spin quartet/triplet, one may discriminate three different transient spin-oscillation types (see Fig.~\ref{sw_types_by_local_states}). Let us focus on the states with extremal $S_z$ to make it obvious. For $N=2.4$, the spin quartet has a higher maximal TD occupation as the spin triplet. As the former has only finite $S_z$ projections, it is more susceptible to net spin polarization since the triplet has one nonmagnetic state ($S_z=0$). So the dynamic change in occupation amplitude triggered by magnetic fluctuations has to be higher for a dominant spin quartet than for the spin triplet. Thus, a quench strength $U_f-$$U_i=0.4$ is sufficient for $N=2.4$, instead for $N=2.8$ an amount
 $U_f-$$U_i=0.7$ is necessary to render the system susceptible to these transient states. The near-degenerate case at $N=2.6$ demands an even higher quench strength. 

\textit{Summary\label{sec:dis}}.---
The connection between spin orientation on the local-correlation timescale and electron-electron interactions for systems with half-filled or hole-doped $t_{2g}$ shell has been investigated. Utilized interaction parameters, magnetic ground states, and excited states (AFM-to-PM, PM-to-AFM) may be directly related to concrete materials cases. Exciting an AFM ground state leads to a nonequilibrium AFM-PM transition for all considered fillings. Furthermore, a broad region of longitudinal large-amplitude spin oscillations emerges at high hole doping. It narrows towards moderate hole doping near the equilibrium AFM-PM phase boundary. These spin oscillations display here a rather robust behavior in time and future developments of more general thermalization schemes beyond TD-SBMFT have to be invoked to investigate their stability. Note that such schemes are at presence still inapplicable to demanding multiorbital Hubbard models.

Transient versions of those spin oscillations appear when exciting a PM state near the equilibrium PM-AFM phase boundary, without a simplistic filling dependence. In the noninteracting ground-state limit for small hole doping, the connection between transient spin oscillation and the equilibrium AFM-PM phase boundary is obvious. The doping-dependent occupation hierarchy of a local spin quartet and triplet within the correlated metal decides between the different nonequilibrium magnetic responses. Note that relevant transversal spin dynamics is expected at lower energy transfers than generally studied in here. Albeit fully accessible within TD-SBMFT, additional symmetry-breaking mechanisms have to be permitted. 
\begin{acknowledgments}
We thank A. I. Lichtenstein for useful discussions. This work has been supported by the DFG cluster of excellence ``The Hamburg Centre
for Ultrafast Imaging'' as well as the DFG-SFB925. 
Computations were performed at the North-German Supercomputing
Alliance (HLRN) under Grant No. hhp00031.
\end{acknowledgments}
\appendix
\section{APPENDIX A: RESULTS FROM DYNAMICAL MEAN-FIELD THEORY}
The dynamical mean-field theory (DMFT) calculations with an continuous-time quantum Monte-Carlo (CT-QMC) impurity solver~\cite{gull_continuous-time_2011,boehnke_orthogonal_2011,parcollet_triqs_????} are performed at $\beta t_h=50$ and $q=0.167$. This allows for a comparison with similar earlier equilibrium phase investigations~\cite{chan_magnetism_2009}. Note that there a Bethe lattice (infinite coordination number and nonloop topology) is used in contrast to our 3D simple cubic dispersion. Within our setup we are able to stabilize AFM order in the hole-doped regime (see Tab. \ref{tab_dmft_results}). We use $U$=15$t_h$, which equals 2.5$D$ as the band width is 12$t_h$. 
\begin{table}[b]
\begin{tabular}{  c | c | c | c | c }
  $N$ & 3.0 & 2.9 & 2.8 & 2.7 \\
  \hline
  $S_z$ & -1.45 & -1.24 & -0.94 & -0.12 \\
  \hline  
\end{tabular}
\caption{Total $S_z$ (summed over orbitals) for appropriate filling $N$ in a 3D simple cubic lattice with $\beta t_h=50$, $q=0.167$ and $U=2.5D$ derived by DMFT using an CT-QMC impurity solver~\cite{gull_continuous-time_2011,boehnke_orthogonal_2011,parcollet_triqs_????}.}
\label{tab_dmft_results}
\end{table} %
\section{APPENDIX B: TOTAL ENERGY CONSIDERATIONS}
In this section, we want to take a closer look at the interplay between the initial total energy $E_i^{tot}$:=$\matrixelement{i}{H\left(U_i\right)}{i}$ and the final total energy one time step after the quench $E_f^{tot}$ in view of the obtained physics. The time evolved parameters (slave bosons $\phi$ and eigenvectors $\nu_a^\bk$) have only acquired changes way below accuracy in the first time step. So, $E_f^{tot}$ is a good approximation to $\matrixelement{i}{H\left(U_f\right)}{i}$, namely, the total energy of the Hamiltonian after the quench in the initial state. 

Note that when we quench the Coulomb interaction $\ham^{\rm loc}_i \neq \ham^{\rm loc}_f$, the potential energy changes abruptly $E_f^{pot}\neq E_i^{pot}$. In contrast, the kinetic energy is time dependent via the renormalization matrices $R$, which are a functional of the time-dependent slave-bosons, and evolves from the initial value $E_i^{kin}$ \cite{lechermann_rotationally_2007}. This means that one time step after the quench $E_f^{kin}$ has acquired only changes way below accuracy leading to $E_f^{kin}=E_i^{kin}$. Note that these considerations are only valid one time step after the quench and that the total energy remains conserved during time evolution after the quench. 
\subsubsection{1. AFM $\rightarrow$ PM}
\begin{figure}[t]
 \centering
 \includegraphics*[width=\columnwidth]{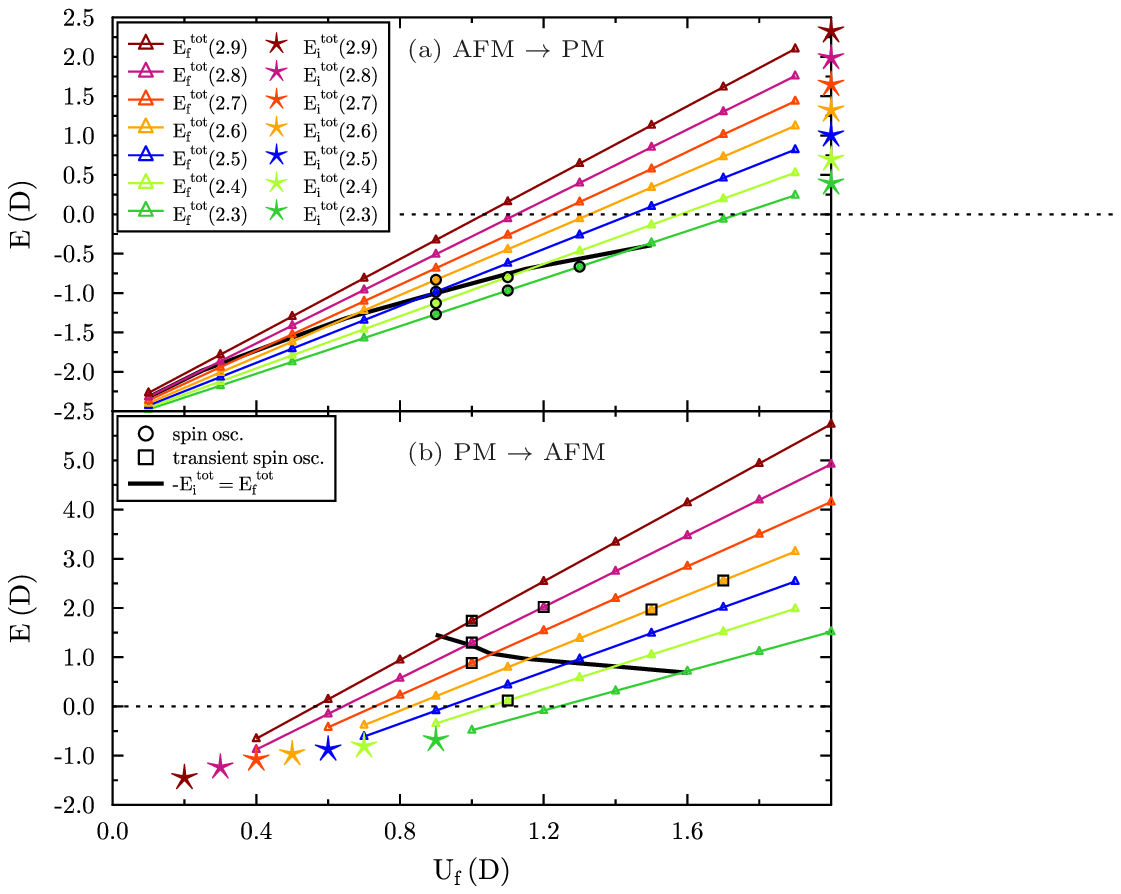}
   \caption{(Color online) $E_i^{tot} (N)$ and $E_f^{tot} (N)$ dependent on filling $N$ and final interaction value $U_f$ with $q=0.2$. For $E_f^{tot} (N) < -E_i^{tot} (N)$ spin oscillations set in as long as AFM fluctuations are present in AFM$\rightarrow$PM case and for $E_f^{tot} (N) \approx -E_i^{tot} (N)$ transient spin oscillations for $N=$2.7-2.9 in PM$\rightarrow$AFM case.}
   \label{totalenergies_filling}
 \end{figure}

First, let us inspect the evolution of both total energies with $U_f$ and filling for the quench scenario AFM$\rightarrow$PM (Fig. \ref{totalenergies_filling}a).
$E_f^{tot}$ exhibits a linear dependence on $U_f$, where the slope of the curve decreases with lower fillings $N$. Note that $E_i^{tot}=E_f^{tot}$($U_f=2.0$) holds. Due to the setup of the quench ($U_f < U_i$) the final total energy is always lower than the initial. A peculiarity is arising as spin oscillations occur, when $E_f^{tot} < -E_i^{tot}$ is reached and an AFM magnetic response is still present (beside the case $N=2.6$). This is not surprising as the kinetic energy remains unchanged $E_i^{kin}=E_f^{kin}$ but the potential energy is lowered by decreasing $U_f$ leading to phase instability. This behavior appears reminiscent of physics contained in the virial theorem, which is, however, not strictly applicable to Hubbard models \cite{capelle_dimensional_2006}. Nonetheless, it is intuitive to assume that a strongly lowered potential energy with unchanged kinetic energy eventually drives the system towards instabilities.

We investigate a straightforward correspondence between spin-oscillation frequencies and total energies (see Tab. \ref{tab_AFM_PM_spin_osc_freq}). This could prove a simple picture describing these spin oscillation in terms of energy scales introduced by both of the total energies. We extract the frequencies using a nonuniform Fourier-transform scheme~\cite{greengard_accelerating_2004,lee_type_2005}, where the frequency resolution is limited by our maximum time of 250$\frac{1}{D}$.
\begin{table}[b]
\begin{tabular}{  l | c  c  c  c }
$\mathbf{\left(N,U_i\right)}=$ &   &   &   &        \\
  $\mathbf{\left(2.3,2.0\right)}$ & $\omega_{osc} $ [$D$]  & $T_{osc}$ [fs] & $E_i^{tot}$ [$D$]  & $E_f^{tot}$ [$D$] \\ 
  \hline
  $U_f=1.3$ & 0.10$\pm$0.03   & 41$\pm$10  & 0.39 & -0.67 \\
  \hline
  $U_f=1.1$ & 0.15$\pm$0.03    & 27$\pm$5  & 0.39 & -0.97 \\
  \hline
  $U_f=0.9$ & 0.20$\pm$0.03   & 21$\pm$3  & 0.39 & -1.27 \\
  \hline
  $\mathbf{\left(2.4,2.0\right)}$ &   &   &  &    \\ 
  \hline
  $U_f=1.1$ & 0.13$\pm$0.03    & 33$\pm$7  & 0.69 & -0.80 \\
  \hline
  $U_f=0.9$ & 0.20$\pm$0.03    & 21$\pm$3  & 0.69 & -1.13 \\
  \hline
  $\mathbf{\left(2.5,2.0\right)}$ &   &   &  &    \\
  \hline
  $U_f=0.9$ & 0.18$\pm$0.03    & 24$\pm$3  & 1.00 & -0.98 \\
  $\mathbf{\left(2.6,2.0\right)}$ &   &   &  &    \\
  \hline
  $U_f=0.9$ & 0.13$\pm$0.03    & 33$\pm$7  & 1.32 & -0.83 \\
\end{tabular}
\caption{Spin oscillation frequencies and their period compared to initial and final total energy in the AFM$\rightarrow$PM quench scenario. Frequencies calculated by nonuniform Fourier transform \cite{greengard_accelerating_2004,lee_type_2005}.}
\label{tab_AFM_PM_spin_osc_freq}
\end{table} %
All spin oscillations lie between 0.10 $\pm$ 0.03 and 0.20 $\pm$ 0.03, which equals 18 to 51 fs (as we are using natural units). This corresponds to frequencies in the THz regime. The frequency decreases linearly with increasing $E_f^{tot}$ at $N=2.3$ with $\omega_{osc}/E_f^{tot} \sim -0.15$. But, already at $N=2.4$ no linear behavior can be derived. Furthermore, the frequency decreases with increasing filling ($E_i^{tot}$) at constant $U_f$. Here again, $\omega_{osc}/E_f^{tot}$ or $\omega_{osc}/(E_f^{tot}-E_i^{tot})$ behave nonlinearly. To conclude, the spin oscillations cannot be explained solely by linear behavior upon $E_f^{tot}$ or $E_f^{tot}-E_i^{tot}$.      
\subsubsection{2. PM $\rightarrow$ AFM}
Let us now turn to the case of quenches from the PM phases. Looking at Fig.~\ref{totalenergies_filling}b, $E_f^{tot}$ exhibits again a linear dependence on $U_f$, where the slope of the curve decreases with lower fillings. As $U_f$ is higher than $U_i$, $E_f^{tot} > E_i^{tot}$ holds in this case. For fillings close to half-filling ($N \in [2.7,2.9]$) transient spin oscillations near $E_f^{tot} = -E_i^{tot}$ are observed, indicating like in the other quench case a phase instability upon increasing the potential energy above a critical value.
Again, linking the occurring frequencies (transient and oscillatory) as well as the transient length to $E_i^{tot}$ and $E_f^{tot}$ provides additional insight into the importance of these energy scales. Transient spin oscillations are evolving into stable AFM oscillations (see Fig. 6 main text). The number of periods of both oscillations is in most cases not sufficient to get an accurate Fourier transform, so we cannot employ a Fourier transform here. Instead, the frequencies will be derived by inspecting the raw time-dependent magnetization data  $m\left(t\right)$ and counting the number of periods (Tab. \ref{tab_PM_AFM_transient}, \ref{tab_PM_AFM_osc} and \ref{tab_PM_AFM_initial0}). Error estimation is done here by considering the largest deviation of $m(t)$ from a sine function ($A\,\sin\left(\omega(x-B)\right)$) with the stated frequency.
\begin{table}[t]
\begin{tabular}{  l | c  c  c  c  }
$\mathbf{\left(N,U_i\right)}=$ &   &   &   &         \\	
  $\mathbf{\left(2.4,0.7\right)}$ & $\omega_{tr} [$D$] $ & $L_{tr}$ [fs]   & $E_i^{tot}$ [$D$]  & $E_f^{tot}$ [$D$] \\ 
  \hline
  $U_f=1.1$ & 0.05$\pm$0.00   & 77$\pm$3 & -0.81  & 0.12 \\
  \hline
  $\mathbf{\left(2.6,0.5\right)}$ &   &   &   &        \\ 
  $U_f=1.5$ & 0.04$\pm$0.00     & 98$\pm$4  & -0.97  & 1.97 \\
  \hline
  $U_f=1.7$ & 0.02$\pm$0.00    & 92$\pm$7   & -0.97  & 2.56 \\
  \hline 
  $\mathbf{\left(2.7,0.4\right)}$ &   &   &   &       \\
  $U_f=1.0$ & 0.08$\pm$0.00     & 25$\pm$1   & -1.08  & 0.88 \\
  \hline
  $\mathbf{\left(2.8,0.3\right)}$ &   &   &   &        \\ 
  $U_f=1.0$ & 0.10$\pm$0.01     & 60$\pm$8  & -1.24  & 1.30 \\
  \hline
  $U_f=1.2$ & 0.07$\pm$0.00   & 64$\pm$3   & -1.24  & 2.02 \\
  \hline 
  $\mathbf{\left(2.9,0.2\right)}$ &   &   &   &       \\ 
  $U_f=1.0$ & 0.11$\pm$0.00    & 19$\pm$1  & -1.46  & 1.74 \\
    \hline
\end{tabular}
\caption{Transient AFM spin oscillation frequencies and transient length  compared to initial and final total energy in the PM$\rightarrow$AFM quench scenario. Frequency derived by inspecting raw data $m\left(t\right)$.}
\label{tab_PM_AFM_transient}
\end{table} %
\begin{table}[h]
\begin{tabular}{  l | c  c  c  c  }
$\mathbf{\left(N,U_i\right)}=$ &   &   &   &         \\	
  $\mathbf{\left(2.4,0.7\right)}$ & $\omega_{osc}$ [$D$] & $T_{osc}$ [fs] & $E_i^{tot}$ [$D$] & $E_f^{tot}$ [$D$] \\
  \hline 
  $U_f=1.1$ & 0.11$\pm$0.02 & 36$\pm$8 & -0.81  & 0.12 \\
  \hline
  $\mathbf{\left(2.6,0.5\right)}$ &   &   &   &        \\ 
  $U_f=1.5$ & - & - & -0.97  & 1.97 \\
  \hline
  $U_f=1.7$ & - & - & -0.97  & 2.56 \\
  \hline 
  $\mathbf{\left(2.7,0.4\right)}$ &   &   &   &       \\
  $U_f=1.0$ & 0.22$\pm$0.09 & 19$\pm$7 & -1.08  & 0.88 \\
  \hline
  $\mathbf{\left(2.8,0.3\right)}$ &   &   &   &        \\ 
  $U_f=1.0$ & 0.23$\pm$0.07 & 18$\pm$5 & -1.24  & 1.30 \\
  \hline
  $U_f=1.2$ & 0.13$\pm$0.03 & 32$\pm$7 & -1.24  & 2.02 \\
  \hline 
  $\mathbf{\left(2.9,0.2\right)}$ &   &   &   &      \\ 
  $U_f=1.0$ & 0.27$\pm$0.08 & 15$\pm$5 & -1.46  & 1.74 \\
    \hline
\end{tabular}
\caption{Stable AFM spin oscillation frequencies and period compared to initial and final total energy in the PM$\rightarrow$AFM quench scenario. Frequency derived by inspecting raw data $m\left(t\right)$. - indicates that frequency could not be measured due to too long transient length.}
\label{tab_PM_AFM_osc}
\end{table} %

The transient spin oscillations have in most cases much lower frequencies than the following stable counterpart (compare Tab. \ref{tab_PM_AFM_transient} and \ref{tab_PM_AFM_osc}). The transient length is varying between 18 and 102 fs (if one does not respect the defined transient oscillation types of Fig. 7 main text). However, there seems to be a link to these different types as it is easier to spin polarize a dominant spin triplet ($N \in$ [2.7,2.9]) as a roughly equal populated spin triplet and quartet ($N=$2.6) leading to much lower transient lengths in the former case. For the same initial conditions, the transient lengths depend differently on $E_f^{tot}$ according to the spin oscillation type. For type 3 ($N=2.7$ to $N=2.9$) there is a proportional behavior and for type 2 an antiproportional behavior. The stable oscillations are given by periods between 12 and 44 fs, so the spin oscillation period region is shifted to lower values compared to the AFM$\rightarrow$PM case. Furthermore, the qualitative behavior of the stable oscillations with filling is different. Increasing the filling (increasing $E_f^{tot}$) at constant $U_f=1.0$ increases $\omega_{osc}$ instead of decreasing it. To clarify the influence of different initial states onto transient and oscillatory behavior, let us look at Tab. \ref{tab_PM_AFM_initial0}.
\begin{table}[t]
\begin{tabular}{  l | c  c  c  c c }
$\mathbf{\left(N,U_i\right)}=$ &   &   &   &  &       \\	
  $\mathbf{\left(2.6,0.0\right)}$ & $\omega_{tr} $ & $L_{tr}$ [fs]  & $\omega_{osc}$ & $E_i^{tot}$   & $E_f^{tot}$ \\ 
  $U_f=0.8$ & 0.04$\pm$0.00   & 52$\pm$5   & 0.12$\pm$0.02  & -1.90  & 0.80 \\
  \hline
  $\mathbf{\left(2.8,0.0\right)}$ &   &   &   &  &      \\ 
  $U_f=0.6$ & 0.07$\pm$0.01     & 140$\pm$23 & - & -1.95  & 0.40 \\
  \hline
  $\mathbf{\left(2.9,0.0\right)}$ &   &   &   &  &     \\
  $U_f=0.6$ & 0.05$\pm$0.01     & 181$\pm$20  & - & -1.98  & 0.54 \\
  \hline
\end{tabular}
\caption{Transient and stable AFM spin oscillation frequencies and transient length compared to initial and final total energy in the PM$\rightarrow$AFM quench scenario. Initial state is not interacting ($U_i$=0). Frequency derived by inspecting raw data $m\left(t\right)$. Unit is $D$ if not mentioned. '-' indicates that frequency could not be measured due to too long transient length.}
\label{tab_PM_AFM_initial0}
\end{table} %
Starting from an initial noninteracting state the transient behavior moves to quenched interaction values right on top of the equilibrium AFM-PM phase boundary. This reveals an intricate connection between this phase boundary and the transient fluctuations. It is not possible to estimate $\omega_{osc}$ in two of the three cases here, as the number of periods is too low. The transient length lies between 50 and 181 fs. Comparing these to Tab. \ref{tab_PM_AFM_transient}, there are drastic changes for $N=2.6$ and $N=2.9$ leading to an increased transient length moving towards half-filling. This is the opposite behavior for $N=2.8$ to $N=2.9$ and the case $U_i \neq 0$, which shows that initial correlations and generally speaking the initial state have a high influence onto the transient length. 
\newpage
 \bibliographystyle{apsrev4-1}
 \bibliography{Paper2}
 \newpage
\end{document}